# Design and Simulation of Vehicle Motion Tracking System using a Youla Controller Output Observation System

**Authors:**
**Rongfei Li (corresponding author)**
University of California, Davis, California
Davis, California 95618, Email: rfli@ucdavis.edu

**Francis F. Assadian**
University of California, Davis, California

**Iman Soltani**
University of California, Davis, California

**Author Contributions:** Methodology, R.L.; Investigation, R.L.; Writing—original draft, R.L.; Writing—review & editing, F.A, I.S. All authors have read and agreed to the published version of the manuscript.

## Statements and declarations

**Ethical considerations:** Not applicable

**Consent to participate:** Not applicable

**Consent for publication:** Not applicable

**Declaration of conflicting interests:** The author(s) declared no potential conflicts of interest with respect to the research, authorship, and/or publication of this article.

**Funding:** This research received no external funding.

**Data Availability Statement:** Data is contained within the article

*Abstract*—*This paper presents a novel linear robust Youla controller output observation system for tracking vehicle motion trajectories using a simple nonlinear kinematic vehicle model, supplemented with positional data from a radar sensor. The proposed system operates across the full vehicle trajectory range with only three linear observers, improving upon previous methods that required four nonlinear observers [1].15 To ensure smooth transitions between Youla controllers and observers, a switching technique is introduced, preventing bumps during controller changes. The proposed observer system is evaluated through simulations, demonstrating accurate and robust estimation of longitudinal and lateral positions, vehicle orientation, and velocity from sensor measurements during various standard driving maneuvers. Results are provided for different driving scenarios, including lane changes and intersection crossings, where significant changes in vehicle orientation occur. The novelty of this work lies in the first application of a Youla controller output observer for vehicle tracking estimation.*

*Index Terms*— *Nonlinear system, Youla controller output observer, MIMO system, Robustness, Vehicle tracking, Hybrid system.*

## 1 Introduction

Vehicle motion detection and tracking have important applications in both civilian and military domains, including urban traffic planning, highway surveillance, and management systems [2]. Traditionally, researchers have explored tracking problems within the field of image processing. Numerous vehicle-tracking methods have demonstrated the ability to detect vehicles and estimate their speeds using visual-based surveillance systems, with solutions primarily relying on image processing techniques [3–6].

In recent years, significant efforts have been directed toward autonomous driving research. Vehicle motion tracking is a crucial challenge for autonomous systems, particularly in areas such as collision avoidance and adaptive cruise control (ACC) [7–8]. These systems commonly employ onboard radar or LiDAR sensors to measure vehicle-to-vehicle (V2V) distances and azimuth angles [7–8].

Interacting Multiple Model (IMM) filters have been widely used for vehicle trajectory estimation since 1970s [9]. The IMM framework allows the simultaneous operation of multiple driving models within a Bayesian framework, with the algorithm dynamically switching between models based on their updated probabilities [10]. Common models employed in IMM filters include the "straight-line driving" model and the "constant turn-rate" model, each suited for specific driving scenarios [1].

Among multiple-model adaptive estimators (MMAE), the Interacting Multiple Model (IMM) approach is particularly computationally efficient, especially for target tracking applications [11–12]. The IMM leverages a bank of filters and models to estimate system states based on the probability associated with each model [13]. It has been implemented with various Kalman filtering techniques, including the linear Kalman filter (KF), extended KF, and unscented KF (UKF) [14].

Recently, neural networks have gained significant popularity across various applications, including vehicle tracking [15-23]. Deep learning techniques, particularly Convolutional Neural Networks (CNNs) [15-18], Recurrent Neural Networks (RNNs) [19-20], and Graph Neural Networks (GNNs) [21-23], dominate current research in this domain. These methods excel at handling complex, non-linear dynamic systems and seamlessly leveraging large datasets. As a result, they are especially effective in analyzing crowded scenarios, such as intersections or merging traffic, where multiple interacting agents exhibit unpredictable behaviors.

However, neural network-based approaches face challenges in scenarios requiring rapid decision-making on the road. Large neural network models with many layers demand significant computational resources, often leading to latency issues that make them unsuitable for real-time applications. Consequently, traditional Interacting Multiple Model (IMM) approaches continue to attract attention in recent research. IMM methods are widely favored in autonomous vehicle applications due to their computational efficiency and adaptability to sudden maneuver changes [24-28]. For example, Anusha et al. [24] demonstrated the application of IMM filters in automotive radar systems for pre-crash scenarios, highlighting their ability to handle rapid motion mode changes within fractions of a second. Similarly, Kämpchen et al. [25] analyzed stop-and-go traffic situations, systematically parameterizing the IMM method based on traffic statistics and validating its performance using real sensor data. Seokwon et al. [26] addressed the challenge of multi-target tracking with a multirotor at long distances, proposing an IMM estimator combined with directional track-to-track association to efficiently manage various maneuvers. These examples underscore the continued relevance of IMM-based approaches in real-time, resource-constrained environments where computational efficiency and adaptability are critical.

Recent research on IMM-based vehicle tracking has focused on improving both accuracy and adaptability. Adaptive IMM algorithms have been developed to adjust model noise variance and Markov matrices dynamically, enhancing tracking precision and real-time performance [29]. Integrating IMM filters with GPS and in-vehicle sensors has further improved accuracy and reliability across various driving conditions [30]. Moreover, the combination of particle filters with IMM algorithms has yielded significant error reduction in cooperative vehicle tracking through V2V communication and GNSS technology [31]. Additionally, IMM frameworks incorporating environment interaction models—accounting for vehicle interdependence—have enhanced the estimation of both lateral and longitudinal motion [32]. These advancements in IMM-based vehicle tracking methods have collectively improved accuracy, adaptability, and reliability in diverse driving scenarios, facilitating the development of autonomous driving systems and smart city applications.

However, despite the improvements in accuracy, the aforementioned approaches rely on multiple sensors or complex online adaptive algorithms. In real-world autonomous vehicle scenarios, one of the biggest challenges is managing computational complexity, as processors must handle multiple tasks simultaneously. These tasks include cruise control to maintain a set distance from the vehicle ahead [33], emergency braking to prevent collisions [34], and blind spot detection [35], etc. Therefore, real-time applications require each task—such as vehicle tracking—to be as efficient as possible in order to minimize computational burden [36–37].

In [1], a single nonlinear kinematic model was employed to represent all possible vehicle motions in urban traffic scenarios.

Using Lyapunov techniques, the paper developed a stable nonlinear observer with four constant gains for state estimation. It also introduced an algorithm to switch between gains across different operating ranges. The proposed observation system, utilizing a single radar sensor and a simple kinematic model, demonstrated good tracking performance in field experiments. Building on this framework, we aim to develop a new observation system with improved performance.

In another work [38], the authors developed a linear Youla Controller Output Observation (YCOO) system based on a nonlinear vehicle dynamic model. The YCOO system is derived from the Youla parametrization technique [39], which is well-suited for handling coupled MIMO (Multiple-Input Multiple-Output) nonlinear systems. After linearization, Youla parametrization decouples MIMO systems into multiple SISO (Single-Input Single-Output) systems using the Smith-McMillan approach. The desired SISO closed-loop system is determined by computing the bandwidth of its associated closed-loop transfer function. Notably, the Youla parametrization technique also preserves closed-loop robustness.

In this paper, the nonlinear system is described through several linear systems via linear approximations at different operating points. A specific linear Youla observer is designed for each of these linear systems. The study shows that only three observers are sufficient to cover the full operational range of vehicle trajectories. The novelties of this paper can be summarized as follows: 1. apply, for the first time, Youla controller output observer for vehicle tracking estimation and 2. compare the performance and robustness of the YCOO system against the previously developed nonlinear observation system in [1]. The YCOO system designed in this study significantly reduces computational burden by using linear observers. Additionally, the results demonstrate improvements in sensor noise reduction and robustness against model parameter variations, highlighting the potential of a YCOO system for more accurate vehicle tracking applications.

This article is organized as follows. In the Background section, we introduce the nonlinear vehicle model used for tracking and provide a brief overview of the nonlinear observer for completeness [1]. In the Methods and Design section, we detail the design of the Youla controller output observers and the switching algorithm that ensures smooth transitions between them. In the Results and Discussion section, we present simulation results comparing the performance and robustness of the YCOO system and the nonlinear observation system across several real-world urban traffic scenarios.

## 2 Background

### 2.1 Vehicle Motion Model

The vehicle model presented in this section is based on the same model described in [1] and is included here for completeness, as shown in Figure 1. Previous approaches using IMM filters typically separate longitudinal and lateral motions into two distinct models [9]. However, the model discussed here integrates both straight-line and turning motions into a unified framework. When planar motion of the vehicle is considered, vehicle motion can be described by four states: X, Y, ψ, V as illustrated in Figure 1. X and Y denote the longitudinal and lateral positions, respectively, while ψ represents the orientation angle of the vehicle relative to the X axis, and V is the vehicle's speed. The positions X and Y are directly measured from the sensor while ψ, V must be estimated from designed observers.

The kinematic equations governing the motion of the vehicle are provided below. These equations describe the relationship between the vehicle's position, orientation, and velocity in a planar environment:

$$\dot{X} = V \cos(\psi + \beta) \tag{1}$$

$$\dot{Y} = V \sin(\psi + \beta) \tag{2}$$

$$\dot{V} = a \tag{3}$$

$$\dot{\psi} = \frac{V\cos(\beta)}{l_f + l_r} \tan(\delta_f) \tag{4}$$

$$\beta = tan^{-1}(\frac{l_r \tan(\delta_f)}{l_f + l_r}) \tag{5}$$

The parameters $l_f$ and $l_r$ are the distances from the center of mass to the front and rear vehicle axles, respectively. The steering angle of the front wheels is denoted by $\delta_f$ while $a$ represents the longitudinal acceleration of the vehicle. Both $\delta_f$ and $a$ are considered inputs to the system, assumed to be either constant or changing slowly over time. The slip angle, denoted as $\beta$, can be calculated using parameters $l_f$, $l_r$ and the input $\delta_f$. These input values are unknown and must be estimated using the developed observers. For further details regarding this vehicle motion model, please refer to [40].

The sum $l_f + l_r$ (wheelbase) is assumed to be 2.8 m for this vehicle model. The actual wheelbase may vary from this value by approximately 10% to 15%. The impact of this variation on the robustness of the system is discussed in the Robustness Analysis section.

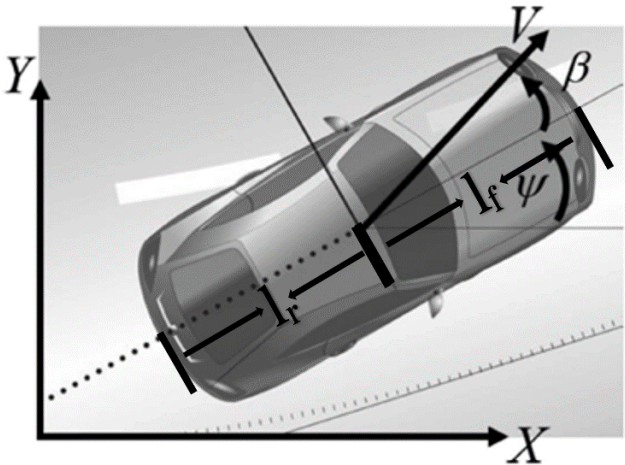

Figure 1. Vehicle motion model [40].

A class of nonlinear systems be represented by the following equations:

$$\dot{s} = f(s, u) \tag{6}$$

$$y = Cs \tag{7}$$

where $s$ is the state vector, $u$ is the input vector and $y$ is the output vector. The matrix $C$ is the output matrix that relates the states to the output. The function $f(s, u)$ is the nonlinear differentiable function that maps states and inputs to the state derivatives.

Associated with the vehicle motion model defined above are the following state and input vectors, respectively:

$$s = [X\ Y\ V\ \psi]^T \tag{8}$$

$$u = [\delta_f, a]^T \tag{9}$$

We can then rewrite equations (1) to (5) as the nonlinear function given in equation (6):

$$f(s,u) = \begin{Bmatrix} V \cos(\psi + \beta) \\ V \sin(\psi + \beta) \\ a \\ \frac{V\cos(\beta)}{l_f + l_r} \tan(\delta_f) \end{Bmatrix} \tag{10}$$

Assuming the longitudinal and lateral positions of the vehicle can be measured by a radar sensor, we can then rewrite (7) as:

$$y = \begin{bmatrix} Y \\ X \end{bmatrix} = Cs \tag{11}$$

where the output matrix C is defined as:

$$C = \begin{bmatrix} 0 & 1 & 0 & 0 \\ 1 & 0 & 0 & 0 \end{bmatrix} \tag{12}$$

## 2.2 A Proposed Algorithm in the Previous Research

The algorithm presented in [1] involves the construction of Luenberger observers with multiple gains to cover the entire operational range. A Luenberger observation system is described by the following equation:

$$\dot{\hat{s}} = f(\hat{s}, u) + L(y - C\hat{s}) \tag{13}$$

where $\hat{s}$ are the estimated states are, and $L$ is the observer gain matrix that needs to be designed to ensure that the estimation error $\tilde{s} = s - \hat{s}$ converges to zero over time.

A theorem introduced in [1] provides a method for determining the observer gain matrix $L$ such that the estimation error from equation (13) converges exponentially to zero. The theorem is stated below:

*Theorem*: Consider the nonlinear system (10) and the observer (13). If there exist matrices $P \geq I$ and $R$ so that the following problem is solvable:

$$\min \; \gamma$$
$$\text{subject to}$$

$$P \geq I \qquad (14)$$

$$\begin{bmatrix} P & R^T \\ R & \gamma I \end{bmatrix} \geq 0 \qquad (15)$$

$$A(s,u)^T P + PA(s,u) - C^T R - R^T C \\ + 2\alpha P \leq 0 \qquad (16)$$

$$\forall u \in u_{grid}, \forall s \in s_{grid} \qquad (17)$$

where $\gamma$ is a parameter to be minimized and $\alpha$ is a positive constant.

The observer gain $L$ is given by:

$$L = P^{-1}R^T \qquad (18)$$

To prevent $L$ from becoming arbitrarily large, an upper bound on the two norms of $L$ must be imposed:

$$\|L\| \leq \sqrt{\gamma} \qquad (19)$$

where $A(s,u) = \frac{\partial f}{\partial s}(s,u)$, which is the Jacobian matrix of the nonlinear function $f(s,u)$. From equation (10), we can find $A(s,u)$ as:

$$A(s,u)$$
$$= \begin{bmatrix} 0 & 0 & \cos(\psi + \beta) & -V\sin(\psi + \beta) \\ 0 & 0 & \sin(\psi + \beta) & V\cos(\psi + \beta) \\ 0 & 0 & 0 & 0 \\ 0 & 0 & \dfrac{\cos(\beta)}{l_f + l_r}\tan(\delta_f) & 0 \end{bmatrix} \qquad (20)$$

It is important to note that the inputs $u$ and states $S$ can vary infinitely within each bound, leading to an infinite number of Linear Matrix Inequalities (LMIs) as indicated in (16). To reduce the number of LMIs, we can grid the inputs and states into finite subsets. The variables $u_{grid}$ and $s_{grid}$ define these gridded subsets of the infinitely varying inputs and states within their boundaries. However, these subsets must be chosen to be sufficiently dense to satisfy stability conditions.

In [1], the boundaries for states and inputs were specifically assumed for urban traffic scenarios. Similarly, we define the following operating ranges:

$$0m/s \leq V \leq 20m/s \qquad (21)$$

$$-20° \leq \delta_f \leq 20° \qquad (22)$$

$$0 \leq \psi \leq 2\pi \qquad (23)$$

To satisfy the condition in (23), the paper designed four constant gain matrices $L_1$, $L_2$, $L_3$, and $L_4$ that operate within the four different operating ranges of $\psi$ respectively:

(1) $-60° \leq \psi \leq 60°$
(2) $\quad 30° \leq \psi \leq 150°$
(3) $\;120° \leq \psi \leq 240°$
(4) $\;210° \leq \psi \leq 330°$

The observers' gains can be obtained by solving (14) − (19) with the LMI toolbox in MATLAB, so as to obtain a single observer gain valid for all values of $A(s,u)$ in the different operating ranges of $\psi$.

The four gain matrices can be shown below:

$$L_1 = \begin{bmatrix} 623.0134 & -1.0612 \times 10^{-12} \\ 1.3971 \times 10^{-12} & 1069.0838 \\ 5388.3054 & -5.5343 \times 10^{-12} \\ 2.1559 \times 10^{-12} & 1528.5641 \end{bmatrix} \qquad (24)$$

$$L_2 = \begin{bmatrix} 1069.0849 & -6.9367 \times 10^{-10} \\ 8.7541 \times 10^{-11} & 623.0136 \\ 3.8624 \times 10^{-9} & 5388.3040 \\ -1528.5669 & 7.2656 \times 10^{-10} \end{bmatrix} \qquad (25)$$

$$L_1 = \begin{bmatrix} 623.0134 & 5.5367 \times 10^{-12} \\ 2.4022 \times 10^{-12} & 1069.0838 \\ -5388.3054 & -3.3334 \times 10^{-11} \\ -2.0986 \times 10^{-12} & -1528.5641 \end{bmatrix} \qquad (26)$$

$$L_1 = \begin{bmatrix} 1069.0849 & -2.7240 \times 10^{-14} \\ 1.4141 \times 10^{-13} & 623.0136 \\ 7.1552 \times 10^{-12} & -5388.3040 \\ 1528.5669 & -9.2341 \times 10^{-11} \end{bmatrix} \qquad (27)$$

By switching the gains between these regions, the Luenberger observation system is ensured to remain stable in the selected operating ranges. Notably, there is a 30° overlap between adjacent regions. The proof of the theorem and additional details of this proposed algorithm can be found in [1].

# 3 Methods and design

The Youla Controller Output Observation (YCOO) system is a model-based estimation technique that designs observers around an estimation model to achieve accurate and stable state estimation. The goal of this section is to design multiple Youla controller output observers in use of Youla parameterization technique ensuring both stability and robustness of the closed-loop system. Three observers are developed to operate across different ranges. The operating ranges of observers are overlapped, so a switching algorithm can be designed to ensure smooth transfer between those observers.

## 3.1 Operating Point Pairs for System Linearization

An overview of the YCOO system is presented in Figure 2. The vehicle motion model is defined by equations (1) to (5). The outputs of this model consist of the estimated longitudinal and lateral positions of the vehicle, which are compared to the actual vehicle positions obtained through a localization sensor such as GPS, a map-based radar or LiDAR, or a camera sensor. The localization error is then fed into Youla-designed observers, allowing the error to be eliminated over time. The outputs of observers include the estimated front wheel steering angle and acceleration, which serve as inputs to the vehicle motion model.

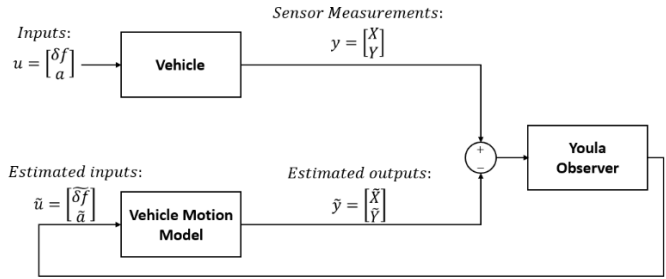

Figure 2. Overview of the YCOO System.

Since the Youla observer can only be designed for linear systems, the nonlinear system described by equation (10) is linearized around different operating point pairs $(s_0, u_0)$ using the first-order Taylor series expansion. After linearization, the nonlinear vehicle tracking system (10) can be represented in a state-space framework:

$$\dot{s} = A(s,u)|_{(s_0,u_0)}(s - s_0) + B(s,u)|_{(s_0,u_0)}(u - u_0) \quad (28)$$

$$y = Cs + Du \quad (29)$$

Where

$$A(s,u)|_{(s_0,u_0)} = \frac{\partial f}{\partial s}(s,u)|_{(s_0,u_0)} \quad (30)$$

$$B(s,u)|_{(s_0,u_0)} = \frac{\partial f}{\partial u}(s,u)|_{(s_0,u_0)} \quad (31)$$

Here, $s$ and $u$ represent the state and actuator input vectors, as defined in equations (8) and (9). In equation (29), the matrix $C$ is given in equation (12), and the feedthrough matrix D is zero.

The matrices $A(s,u)|_{(s_0,u_0)}$ and $B(s,u)|_{(s_0,u_0)}$ at the operating point $(s_0, u_0)$ can be expressed as follows:

$$A(s,u)|_{(s_0,u_0)}$$
$$= \begin{bmatrix} 0 & 0 & cos(\psi_0 + \beta_0) & -V_0 sin(\psi_0 + \beta_0) \\ 0 & 0 & sin(\psi_0 + \beta_0) & V_0 cos(\psi_0 + \beta_0) \\ 0 & 0 & 0 & 0 \\ 0 & 0 & \frac{cos(\beta_0)}{l_f + l_r} tan(\delta_{f_0}) & 0 \end{bmatrix} \quad (32)$$

$$B(s,u)|_{(s_0,u_0)} =$$
$$\begin{bmatrix} -V_0 sin(\psi_0 + \beta_0) \frac{\partial \beta}{\partial \delta_f}|_{(\delta_{f_0}=0)} & 0 \\ V_0 cos(\psi_0 + \beta_0) \frac{\partial \beta}{\partial \delta_f}|_{(\delta_{f_0}=0)} & 0 \\ 0 & 1 \\ \partial(\frac{cos(\beta)}{l_f + l_r} tan(\delta_f))/\partial \delta_f|_{(\delta_{f_0}=0)} & 0 \end{bmatrix} \quad (33)$$

where $A(s,u)|_{(s_0,u_0)}$ and $B(s,u)|_{(s_0,u_0)}$ are functions of states $V$, $\psi$, and the input $\delta_f$ at their operating points $V_0, \psi_0$ and $\delta_{f_0}$. Also,

$$\beta_0 = tan^{-1}(\frac{l_r tan(\delta_{f_0})}{l_f + l_r}) \quad (34)$$

The next step is to determine the values of the operating point pairs $(s_0, u_0)$. Assume the operating ranges are identical to those defined in inequalities (21), (22), and (23).

By examining the nonlinear system in equation (10) and holding all other variables ($\psi$, a, $\delta_f$ and $\beta$) constant, it becomes evident that the vehicle speed $V$ is linearly proportional to the terms involving $V$ in their expressions. Therefore, any value within the operating range of $V$ specified in equation (21) can serve as a valid operating point $V_0$. For simplicity, we select $V_0$ as the average of the upper and lower bounds of this range:

$$V_0 = 10m/s \quad (35)$$

Given the operating range of $\delta_f$ in inequality (22), we can approximate:

$$tan(\delta_f) \cong \delta_f \quad (36)$$

using the first-order Taylor series expansion, which holds for small angles. Therefore, the following relation holds:

$$\frac{l_r tan(\delta_f)}{l_f + l_r} \cong \frac{l_r \delta_f}{l_f + l_r} \quad (37)$$

Assuming the vehicle exhibits understeering behavior, the following condition applies:

$$\frac{l_r}{l_f + l_r} < 0.5 \quad (38)$$

From inequalities (22) and (38), we can establish the range of equation (37) as:

$$\left| \frac{l_r tan(\delta_f)}{l_f + l_r} \right| \cong \left| \frac{l_r \delta_f}{l_f + l_r} \right| < 10° \quad (39)$$

For small values of $\frac{l_r tan(\delta_f)}{l_f + l_r}$, the first-order Taylor series gives

$$tan^{-1}\left(\frac{l_r\,tan(\delta_f)}{l_f+l_r}\right) \cong \frac{l_r\,tan(\delta_f)}{l_f+l_r} \qquad (40)$$

Thus, the slip angle $\beta$\beta$\beta$ in equation (5) can be approximated as:

$$\beta = tan^{-1}\left(\frac{l_r\,tan(\delta_f)}{l_f+l_r}\right)$$
$$\cong \frac{l_r\,tan(\delta_f)}{l_f+l_r} \cong \frac{l_r\delta_f}{l_f+l_r} \qquad (41)$$

From (39) and (41) we can derive the non-strict operating range of $\beta$ as:

$$-10° < \beta < 10° \qquad (42)$$

Since the operating range of $\beta$\beta$\beta$ is small, we can apply the small-angle approximations:

$$cos(\beta) \cong 1 \qquad (43)$$

$$sin(\beta) \cong \beta \qquad (44)$$

As a result, the following simplifications hold:

$$Vcos(\psi + \beta) \cong Vcos(\psi) \qquad (45)$$

$$Vsin(\psi + \beta) \cong Vsin(\psi) \qquad (46)$$

Additionally, the fourth term in equation (10) simplifies as:

$$\frac{Vcos(\beta)}{l_f+l_r}tan(\delta_f) \cong \frac{V}{l_f+l_r}\delta_f \qquad (47)$$

The overall approximate forms of (10) are provided below:

$$f(s,u) \cong \begin{Bmatrix} V\,cos(\psi) \\ V\,sin(\psi) \\ a \\ \frac{V}{l_f+l_r}\delta_f \end{Bmatrix} \qquad (48)$$

Similarly, by examining the nonlinear system in equation (10) and holding all other variables ($V, a$ and $\psi$) constant, it becomes evident that the steering angle, $\delta_f$, is linearly proportional to the fourth term in equation (48). Therefore, any value within the operating range of $\delta_f$ specified in inequality (22) can serve as a valid operating point, $\delta_{f_0}$. For simplicity, we select $\delta_{f_0}$ as the average of the upper and lower bounds of this range:

$$\delta_{f_0} = 0° \qquad (49)$$

To ensure stability and robustness of YCOO, we divide the operating range of $\psi$ into three sub-regions, each with a different average orientation angle $\psi_0$.

## 3.2 Observer Design by Youla Parameterization

As stated in section 3.1, our objective is to design observers that stabilize the closed-loop system. To start, let us assume that

$$\psi_0 = 0° \qquad (50)$$

We initially use this assumption to design an observer based on Youla Parameterization. The next step is to determine the corresponding range of $\psi$ that ensures a stable closed-loop system.

We begin by using the operating point pairs $(s_0, u_0)$ provided in equations (35), (49) and (50) to numerically compute the two Jacobian matrices $A(s, u)$ and $B(s, u)$:

$$A(s,u)\Big|_{\left(\left[V_0, \psi_0, \delta_{f_0}\right] = \left[\frac{10m}{s}, 0°, 0°\right]\right)}$$
$$= \begin{bmatrix} 0 & 0 & 1 & 0 \\ 0 & 0 & 0 & 10 \\ 0 & 0 & 0 & 0 \\ 0 & 0 & 0 & 0 \end{bmatrix} \qquad (51)$$

$$B(s,u)\Big|_{\left(\left[V_0, \psi_0, \delta_{f_0}\right] = \left[\frac{10m}{s}, 0°, 0°\right]\right)}$$
$$= \begin{bmatrix} 0 & 0 \\ 5.1786 & 0 \\ 0 & 1 \\ 3.5714 & 0 \end{bmatrix} \qquad (52)$$

Given the system parameters, the multivariable plant transfer function matrix is derived from the state space representation in equations (28) and (29) It can be expressed as:

$$G_p = \begin{bmatrix} \dfrac{Y(s)}{\delta_f(s)} & \dfrac{X(s)}{\delta_f(s)} \\ \dfrac{Y(s)}{a(s)} & \dfrac{X(s)}{a(s)} \end{bmatrix}$$
$$= \begin{bmatrix} \dfrac{5.179s + 35.71}{s^2} & 0 \\ 0 & \dfrac{1}{s^2} \end{bmatrix} \qquad (53)$$

The first step in designing an observer for this multivariable system using Youla Parameterization is to determine the Smith-McMillan form of the plant [41].

The Smith-McMillan form of the plant $G_p$ can be found as:

$$M_p = U_L G_p U_R \qquad (54)$$

where $U_L$ and $U_r$ are the left and right unimodular matrices, and $M_p$ is the Smith-McMillan form of $G_p$. The unimodular matrices are given as:

$$U_L = \begin{bmatrix} \dfrac{0.1931}{(s + 6.897)} & 0 \\ 0 & (s + 6.897) \end{bmatrix} \qquad (55)$$

$$U_R = \begin{bmatrix} 1 & 0 \\ 0 & 1 \end{bmatrix} \tag{56}$$

Thus, the Smith-McMillan form $M_p$ is:

$$M_p = \begin{bmatrix} M_{p1} & 0 \\ 0 & M_{p2} \end{bmatrix} = \begin{bmatrix} \frac{1}{s^2} & 0 \\ 0 & \frac{(s+6.897)}{s^2} \end{bmatrix} \tag{57}$$

where $M_{p1}$ and $M_{p2}$ represent decoupled plant of $G_p$. Both $M_{p1}$ and $M_{p2}$ have two poles at the origin, making the plant Bounded input Bounded output (BIBO) unstable. To ensure internal stability of the closed-loop system, the following interpolation conditions must be satisfied [42]:

$$M_{T1}(s = 0) = 1, \qquad M_{T2}(s = 0) = 1 \tag{58}$$

$$\frac{dM_{T1}}{ds}\big|_{s=0} = 0, \qquad \frac{dM_{T2}}{ds}\big|_{s=0} = 0 \tag{59}$$

where $M_{T1}$ and $M_{T2}$ represent the decoupled closed loop transfer functions. These functions can be expressed as:

$$M_{T1} = M_{p1}M_{y1}, \qquad M_{T2} = M_{p2}M_{y2} \tag{60}$$

where $M_{y1}$ and $M_{y2}$ are the decoupled Youla transfer functions. These transfer functions are selected to ensure that the interpolation conditions in equations (58) and (59) are satisfied.

The decoupled sensitivity transfer functions $M_{s1}$ and $M_{s2}$ are given by:

$$M_{s1} = 1 - M_{T1}, \quad M_{s2} = 1 - M_{T2} \tag{61}$$

The selected forms of the decoupled closed-loop transfer (details omitted due to space constraints) are

$$M_{T1} = \frac{(3w_1^2 s + w_1^3)}{(s + w_1)^3(0.001s + 1)} \tag{62}$$

$$M_{T2} = \frac{(3w_2^2 s + w_2^3)}{(s + w_2)^3(0.001s + 1)} \tag{63}$$

These transfer functions can be written in a decoupled matrix form as:

$$M_T = \begin{bmatrix} \frac{(3w_1^2 s + w_1^3)}{(s + w_1)^3(0.001s + 1)} & 0 \\ 0 & \frac{(3w_2^2 s + w_2^3)}{(s + w_2)^3(0.001s + 1)} \end{bmatrix} \tag{64}$$

Here, $w_1$ and $w_2$ determine the dominant pole locations of the closed-loop system. The additional poles at $(0.001s + 1)$ is introduced to reduce the magnitude of the closed-loop transfer functions at high frequencies, providing better sensor noise rejection. This pole is selected to be at least twice as larger as the dominant poles $w_1$ and $w_2$, ensuring it does not interfere with the closed-loop system's bandwidth.

It can be verified that our selections for the closed loop transfer functions in (62) and (63) satisfy the constraints given in (58) and (59). We can compute the decoupled Youla transfer functions as follows:

$$M_y = \frac{M_T}{M_p}$$

$$= \begin{bmatrix} \frac{s^2(3w_1^2 s + w_1^3)}{(s + w_1)^3(0.001s + 1)} & 0 \\ 0 & \frac{s^2(3w_2^2 s + w_2^3)}{(s + 6.897)(s + w_2)^3(0.001s + 1)} \end{bmatrix} \tag{65}$$

Through trials and errors, it has been observed that the difference between the parameters $w_1$ and $w_2$ ($w_1 > w_2$), influences the operating range of a single Youla observer. A greater difference expands the operating range. However, $w_1$ cannot be increased beyond a certain limit; otherwise, the bandwidth of the system $M_{T1}$ becomes excessively large, potentially amplifying high-frequency sensor noise. At the same time, $w_2$ cannot be decreased below a threshold, or else the bandwidth of the system $M_{T2}$ becomes too small, causing the observer to respond slowly. After numerous simulations, the optimal combination of $w_1$ and $w_2$ that balances performance and robustness within the observer's operating range is found to be:

$$w_1 = 500 \; rad/s \quad w_2 = 30 rad/s \tag{66}$$

The coupled Youla, closed loop, sensitivity, and observer transfer function matrices are computed as follows:

$$Y = U_R M_y U_L \tag{67}$$

$$T_y = G_p Y \tag{68}$$

$$S_y = 1 - T_y \tag{69}$$

$$G_C = Y S_y^{-1} \tag{70}$$

The frequency responses of $T_y$, $S_y$, $Y$ and $G_C$ for the output-input pairs $(Y(s), \delta_f(s))$, $(X(s), a(s))$ are presented in Figures 3 and 4, respectively. As shown in these figures, the point where $T_y$ and $S_y$ are intersected is the crossover frequency. For the closed loop $T_y = \frac{Y(s)}{\delta_f(s)}$, the bandwidth is around 500rad/s while for the closed loop $T_y = \frac{X(s)}{a(s)}$, the bandwidth is around 30rad/s. These values align with the specifications of the designed system in equation (62). At frequencies below the bandwidth, the gain of $T_y$ is 1 and $S_y$ is 0, ensuring effective target following and robustness against model uncertainties. Above the bandwidth frequency, the gain of $T_y$ falls below 1, allowing the system to reject high-frequency noise. Similarly, the magnitudes of $Y$ and $G_C$ also decline at high frequencies to protect the system from noise.

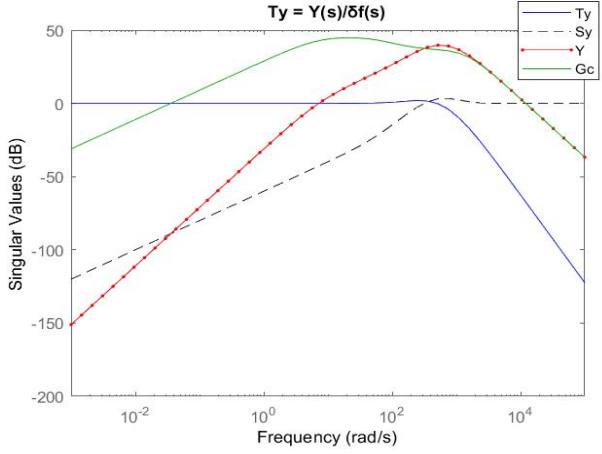

Figure 3. Frequency response of $T_y$, $S_y$, $Y$ and $G_C$ for output-input pair $(Y, \delta_f)$.

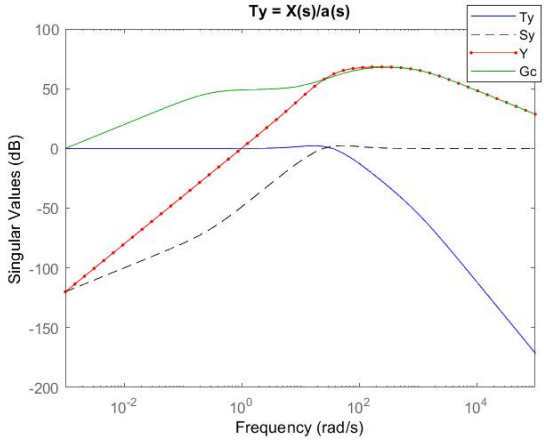

Figure 4. Frequency response of $T_y$, $S_y$, $Y$ and $G_C$ for output-input pair $(X, a)$.

## 3.3 Algorithm for Bumpless Transfer of Observer Gains

In section 3.2, $\psi_0$ is assumed to be $0°$. However, an observer designed under this assumption cannot guarantee stability across all operating ranges of $\psi$. To ensure a stable response that encompasses the entire range of $\psi$, two additional observers are implemented, designed around two other linearized operating points: $\psi_0 = 120°$ and $\psi_0 = 240°$. The linearized plants, $G_p$, and the corresponding observers, $G_c$, for each of the three operating points $\psi_0$ are provided below. The Youla parameterization technique described in Section 3.2 can be applied to design these observers. The plants and observers derived from each linearized operating point $\psi_o$ are summarized below.

With $\psi_o = 0°$:

$$G_{p1} = \begin{bmatrix} \frac{5.179s+35.71}{s^2} & 0 \\ 0 & \frac{1}{s^2} \end{bmatrix} \tag{71}$$

$$G_{c1} =$$

$$\begin{bmatrix} \frac{1.4483*10^8(s+166.7)s}{(s+6.87)(s+59.38)(s^2+2441s+2.105*10^6)} & 0 \\ 0 & \frac{2.7*10^6(s+10)s}{(s+92.65)(s+997.1)(s+0.2923)} \end{bmatrix} \tag{72}$$

With $\psi_o = 120°$:

$$G_{p2} = \begin{bmatrix} \frac{-2.589s-17.86}{s^2} & \frac{-4.485s-30.93}{s^2} \\ \frac{0.866}{s^2} & \frac{-0.5}{s^2} \end{bmatrix} \tag{73}$$

$$G_{c2} = \begin{bmatrix} \frac{-3.5483*10^7(s+166.7)s}{(s+6.897)(s+31.68)(s^2+2018s+1.354*10^6)} \\ \frac{-6.1458*10^7(s+116.7)s}{(s+6.897)(s+31.68)(s^2+2018s+1.354*10^6)} \end{bmatrix}$$

$$\begin{bmatrix} \frac{1.6238*10^{12}(s+8.333)s}{(s+0.5863)(s+80.99)(s+860)(s^2+2133s+1.148*10^6)} \\ \frac{-9.375*10^{11}(s+8.333)s}{(s+80.99)(s+860)(s+0.5863)(s^2+2133s+1.148*10^6)} \end{bmatrix} \tag{74}$$

With $\psi_o = 240°$:

$$G_{p3} = \begin{bmatrix} \frac{-2.589s-17.86}{s^2} & \frac{4.485s+30.93}{s^2} \\ \frac{-0.866}{s^2} & \frac{-0.5}{s^2} \end{bmatrix} \tag{75}$$

$$G_{c3} = \begin{bmatrix} \frac{-3.5483*10^7(s+116.7)s}{(s+6.897)(s+31.68)(s^2+2018s+1.354*10^6)} \\ \frac{-1.6238*10^{12}(s+8.333)s}{(s+0.5863)(s+80.99)(s+860)(s^2+2133s+1.148*10^6)} \end{bmatrix}$$

$$\begin{bmatrix} \frac{6.1458*10^7(s+116.7)s}{(s+6.897)(s+31.68)(s^2+2018s+1.354*10^6)} \\ \frac{-9.375*10^{11}(s+8.333)s}{(s+80.99)(s+860)(s+0.5863)(s^2+2133s+1.148*10^6)} \end{bmatrix} \tag{76}$$

For robust and stable performance, each observer must operate within specific ranges, which are summarized in Table 1.

TABLE I
OPERATION RANGE OF OBSERVERS

| Observer Number | Observer Associated | Linearize point $\psi_0$ | Operation range of $\psi$ |
|---|---|---|---|
| 1 | $G_{c1}$ | $\psi_0 = 0^o$ | [-70$^o$ , 70$^o$] |
| 2 | $G_{c2}$ | $\psi_0 = 120^o$ | [50$^o$ , 190$^o$] |
| 3 | $G_{c3}$ | $\psi_0 = 240^o$ | [170$^o$ , 310$^o$] |

Note that there is a $20^o$ overlap between adjacent operating ranges.

Simulation results indicate that the response becomes unstable when the state $\psi$ deviates from the linearized point $\psi_0$ beyond a certain range. For each observer, the stability operating range is given by $[-70^o + \psi_0, 70^o + \psi_0]$. If the state is in the non-overlapped range, we just utilize the designed observer for that range as other observers result in unstable responses. When the state $\psi$ falls within the overlapping ranges, specifically $\psi \in [50°,70°]$, $\psi \in [170°,190°]$, or $\psi \in [290°,310°]$, either of the two adjacent observers can be considered.

Due to the deviation from the linearized model to the nonlinear model, errors may arise between the estimated states and their actual values. To quantify the performance of each observer at different orientations, we utilize the root mean square of errors ($RMS$) between the estimated and actual values of the state $\psi$. The $P_{residues}$ over a 10-second period for each observer is summarized in Tables 2-4. These tables present the performance of each observer when the orientation angle $\psi$ is within the overlapping range. Note that a smaller $RMS$ value indicates better observer performance in that range.

TABLE II
$RMS$ OF OBSERVERS IN OVERLAPPED RANGE" $\Psi \in [50°,70°]$"

| $\psi$ | $Observer\ 1$ | $Observer\ 2$ |
|---|---|---|
| 50° | 0.0023 | 0.006 |
| 55° | 0.0026 | 0.0055 |
| 60° | 0.0029 | 0.0051 |
| 65° | 0.0031 | 0.0046 |
| 70° | 0.0033 | 0.0041 |

TABLE III
$RMS$ OF OBSERVERS IN OVERLAPPED RANGE" $\Psi \in [170°,190°]$"

| $\psi$ | $Observer\ 2$ | $Observer\ 3$ |
|---|---|---|
| 170° | 0.0041 | 0.006 |
| 175° | 0.0046 | 0.0055 |
| 180° | 0.0051 | 0.0051 |
| 185° | 0.0055 | 0.0046 |
| 190° | 0.006 | 0.0041 |

TABLE IV
$RMS$ OF OBSERVERS IN OVERLAPPED RANGE" $\Psi \in [290°,310°]$"

| $\psi$ | $Observer\ 3$ | $Observer\ 1$ |
|---|---|---|
| 290° | 0.0041 | 0.0033 |
| 295° | 0.0046 | 0.0031 |
| 300° | 0.0051 | 0.0029 |
| 305° | 0.0055 | 0.0026 |
| 310° | 0.006 | 0.0023 |

The shaded area represents the $RMS$ for each observer operating at a specific orientation angle $\psi$ and $RMS$ is calculated as followings:

$$RMS = \sqrt{\frac{1}{N}\sum_{i=1}^{N} R_i^{\,2}} \quad (77)$$

Where $R_i$ is the residue of the $i^{th}$ sample of the signal and $N$ is the number of samples in the simulations.

The units of these values are $degrees$. To estimate $RMS$ for values not listed in the tables, linear interpolation can be employed.

A bump-less transfer algorithm is needed to ensure smooth transitions when switching between observers without sudden overshoots in the estimation outputs. To minimize the overshoot caused by abrupt switching between observers, the outputs of the observers are combined in the overlapping range by applying weights based on each observer's performance (or $RMS$ values) in that range. The weights and the combined observer outputs are computed as follows:

$$W_i = \frac{RMS_j}{RMS_i + RMS_j} \quad (78)$$

$$W_j = \frac{RMS_i}{RMS_i + RMS_j} \quad (79)$$

$$\tilde{u}_c = W_i\tilde{u}_i + W_j\tilde{u}_j \quad (80)$$

where $W_i$ and $W_j$ are the weights for observer $i$ and observer $j$ respectively. In the overlapped ranges, these weights are calculated based on their SSR values from Tables II-IV. The term $\tilde{u}_c$ represents the combined estimated output that is fed into the vehicle model.

Figure 5 illustrates a block diagram of the proposed bump-less transfer algorithm. Here, $\tilde{\psi}$ denotes the estimated orientation angle of the vehicle and the vector $\begin{bmatrix} e_Y \\ e_X \end{bmatrix}$ holds the estimation error for the states $X$ and $Y$. When $\tilde{\psi}$ falls within a non-overlapping range, only one observer's output is utilized, with its weighting value set to one, while the weights of the other observers' outputs are set to zero.

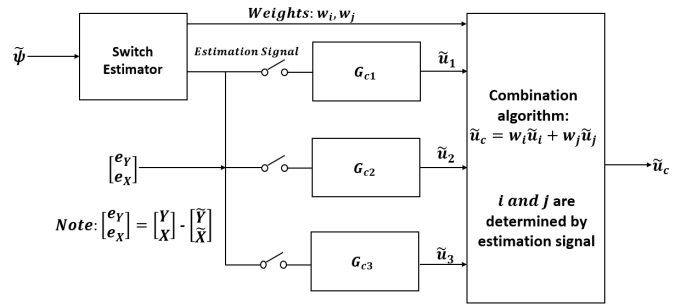

Figure 5. Diagram of bump-less transfer

The stability proof for this bump-less transfer algorithm is provided in [43].

# 4 Results

The performance and robustness of both the nonlinear observation system [1] and the Youla Controller Output Observation (YCOO) system proposed in this work have been evaluated through simulations. The simulations cover a range of real-world vehicle motion scenarios commonly encountered on urban roads and intersections. These scenarios include straight-line driving (Figure 6(a)), lane changes (Figure 6(b)), double lane changes (Figure 6(c)), cross-traffic interactions (Figure 6(d)), and left turns (Figure 6(e)). The range of vehicle speeds across all scenarios is maintained between 5 m/s -15 m/s (or approximately 11 miles/hour - 33 miles/hour) to realistically simulate urban traffic conditions. For the first three scenarios—straight-line driving, lane changes, and double lane changes—the vehicle's orientation angle $\psi$ remains within the range $-20^o \leq \psi \leq 20^o$. Based on the bump-less transfer algorithm discussed previously, only Observer 1 is activated in these scenarios. In the cross-traffic scenario, Observer 3 is exclusively utilized, as the vehicle's orientation angle ($\psi$) is maintained at a constant value of 270° throughout the simulation. Conversely, the left-turn scenario involves more complex maneuvers, requiring the activation of a combination of Observer 2 and Observer 3. Smooth transitions between these observers are implemented to maintain accurate vehicle motion estimation. Figure 6 illustrates all five vehicle maneuvers, assuming a radar sensor is located at the origin of the coordinate system. These scenarios test the ability of the proposed system to adapt to varying vehicle dynamics and motion complexities.

All five scenarios are simulated using both the nonlinear observation system and the YCOO system. To assess their sensor noise rejection capabilities, white noise with a power of $0.01\ m^2$ is introduced into the closed-loop system. This noise directly impacts the radar sensor output estimations, specifically the current X and Y coordinates of the vehicle. Each scenario is simulated with 30 repetitive runs with random generated noises in MATLAB with the same noise power. Estimation results for all states are shown in Figures 7 to 11, with absolute errors between estimated and actual values plotted on the right side of each figure to highlight the differences in performance under noise. Additionally, the acceleration and steering angle values responsible for each maneuver are presented in the bottom plots of each figure, though these values serve only to generate the simulations; in practical applications, they are unknown to the observers. The average (over 30 runs) root mean square (RMS) of errors over the entire simulation period—10 seconds for the first four scenarios and 20 seconds for the left-turn scenario—is compared between the YCOO system and the nonlinear observation system, with the results summarized in Table V. The standard deviations of RMS of errors over 30 runs are also included in Table VI. Additionally, the average frequency of errors for both systems is compared and presented in Table VIII. For the nonlinear observation system, gains and additional parameters used in the simulations are provided in paper [1].

By analyzing the average and standard deviations of the RMS errors, we can statistically compare the denoising performances of the YCOO system and the nonlinear observation system. To assess whether the observed differences are statistically significant, we introduce the p-value [44] as a measure of significance. The hypothesis testing is defined as follows:

- Null Hypothesis ($H_0$): There is no significant difference between the two methods in denoising white noise.
- Alternative Hypothesis ($H_A$) : There is a significant difference between the two methods in denoising white noise.

A significance level (α) of 0.05 is used in this study. This means:

- If the p-value is less than 0.05, we reject $H_0$, indicating a statistically significant difference between the two methods.
- If the p-value is greater than or equal to 0.05, we fail to reject $H_0$, implying that there is insufficient evidence to conclude a significant difference.

The p-values are computed based on the t-distribution and the degrees of freedom, as outlined in [44]. The results for each scenario are summarized in Table VII.

The smallest decimal precision across all states is assumed to be one decimal place. Consequently, tolerance is defined as half of this smallest decimal precision. Specifically, the RMS error tolerance is set to 0.05 m for vehicle trajectory positions, 0.05° for orientation angles, and 0.05 m/s for speeds. This tolerance acts as an evaluation metric to determine whether observers have effectively reduced errors within the acceptable range when incorporating sensor noise or varying model parameters.

In real-world applications, the observer must compute estimated outputs at a specific frequency to ensure accurate and timely results. In this paper, a kinematic vehicle model is used, so the required computation frequency does not need to capture the fast dynamics typically present in more complex systems. However, when radar sensors are used, they have inherent update rates that dictate how often new data is available. To ensure accuracy and minimize latency, the computation frequency of the observer should be at least 10 times the sensor update rate. This ensures that the observer can process data efficiently, maintain synchronization with the sensors, and prevent errors caused by under-sampling or delayed responses.

Delay is another inevitable challenge in real-world observation system applications. It can stem from sensor data processing, communication latency, or computational complexity. A common approach to address constant pure time delays in systems is the Smith Predictor [45]. This method predicts the system's future state using a delay-free model and applies compensation to mitigate the impact of the delay. However, the development and implementation of the Smith Predictor are beyond the scope of this paper and may be explored in future work.

Observed from Figures 7-11, both systems demonstrate strong performance in estimating all states across various vehicle motion scenarios. The absolute error plots indicate that the YCOO system is more effective at filtering out high-frequency noise than the nonlinear observation system, as its error signals are primarily in the lower frequency range. The YCOO system suffers less discrepancy between estimated and actual values of states, particularly in the estimation of orientation angles $\psi$. This effectiveness is attributed to the closed-loop parameter Ty in the Youla parameterization controller design, which remains very small at high frequencies, thereby preventing the excitation of noise, as discussed in Section 3.2.

Table V presents the average RMS values of all states—trajectory, orientation angle, and speed—across all simulation scenarios, providing a quantitative comparison between the YCOO system and the nonlinear observation system. The comparison reveals that the YCOO system significantly reduces the magnitudes of errors in all states, which aligns with the error plots shown in Figures 7–11. Both systems successfully reduce errors within the set tolerance for trajectory positions and speeds. However, the nonlinear system fails to attenuate orientation angle errors within the acceptable range in all traffic scenarios, whereas the YCOO system only fails in the Cross-traffic case.

Table VII presents the p-values for comparing the two methods in state estimation, based on the average RMS values (Table V) and the average standard deviation of errors (Table VI) over 30 simulation runs. With the exception of vehicle trajectory estimation in the straight-line and lane-change scenarios, the p-values in Table VII are below the significance level of 0.05, indicating a statistically significant difference between the two methods in denoising white noise. Since the average RMS in the YCOO system is consistently lower than that in the nonlinear observation system across all states and scenarios, we can conclude that YCOO significantly improves state estimation performance in most scenarios, with a confidence level exceeding 95%.

White noise, characterized by frequencies ranging from low to high with uniform intensity or power, has been introduced into the system. Table VIII details the average error frequencies of all states across all driving scenarios. The results show that the average frequencies of error signals in the YCOO system are significantly lower than those in the nonlinear system, indicating that the YCOO system more effectively attenuates high-frequency components of white noise compared to the nonlinear system.

However, the YCOO system experiences more overshoots during discontinuous state changes, as evident in the speed plots across various scenarios (Figures 7, 9–11). These phenomena are particularly noticeable in the enlarged speed plots within these figures. This discrepancy is attributed to delays in the linear observers of the YCOO system when tracking the nonlinear vehicle models.

Robustness analysis has also been conducted for both observation systems. In Section 2.1, the parameters $l_f$ and $l_r$ of the vehicle model (Equations 1–5) were introduced. The total length $l_f + l_r$ represents the distance from the front to the rear of the vehicle, known as the wheelbase, denoted as $l_t$. The subsequent analysis examines the impact of varying $l_t$ on the estimations of velocity and orientation angle. Figures 12 and 13 illustrate the results of the robustness tests conducted during a double lane change maneuver, without injecting sensor noise into the simulations.

Figure 12 presents the estimated velocities from both observation systems under varying $l_t$. During orientation changes (between 2 and 6 seconds), estimation errors between the model with parameter variations and the nominal model become significant. The YCOO system demonstrates greater resilience to these parameter variations, with its velocity estimation being less affected compared to the nonlinear observation system.

The impact of parameter variations on orientation angle estimation is depicted in Figure 13. The YCOO system exhibits near-perfect robustness to these parameter variations, attributed to its low sensitivity in the low-frequency domain. This sensitivity is quantified by the magnitude of the transfer function matrix Sy, as shown in Figures 3 and 4. Variations in $l_t$ have a more pronounced negative effect on the nonlinear observation system's performance than on the YCOO system.

Table IX summarizes the RMS values for the entire 10-second simulation, with varying $l_t$ for both YCOO and nonlinear observers. he YCOO system successfully handles model variations by keeping errors within the set tolerance, while the nonlinear observation system fails the robustness test when $l_t$ is either increased or decreased. These RMS values further demonstrate that the YCOO system delivers superior robustness against model variations compared to the nonlinear system.

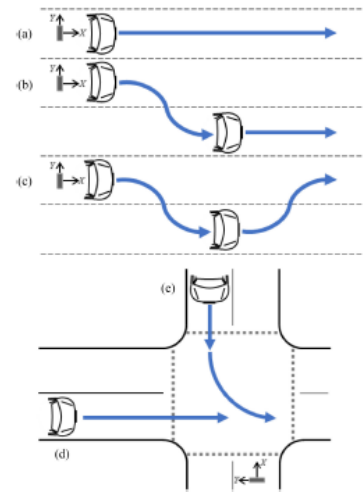

Figure 6. Vehicle motions. (a) Straight line. (b) Lane change. (c) Double lane change. (d) Cross traffic. (e) Left turn. [1]

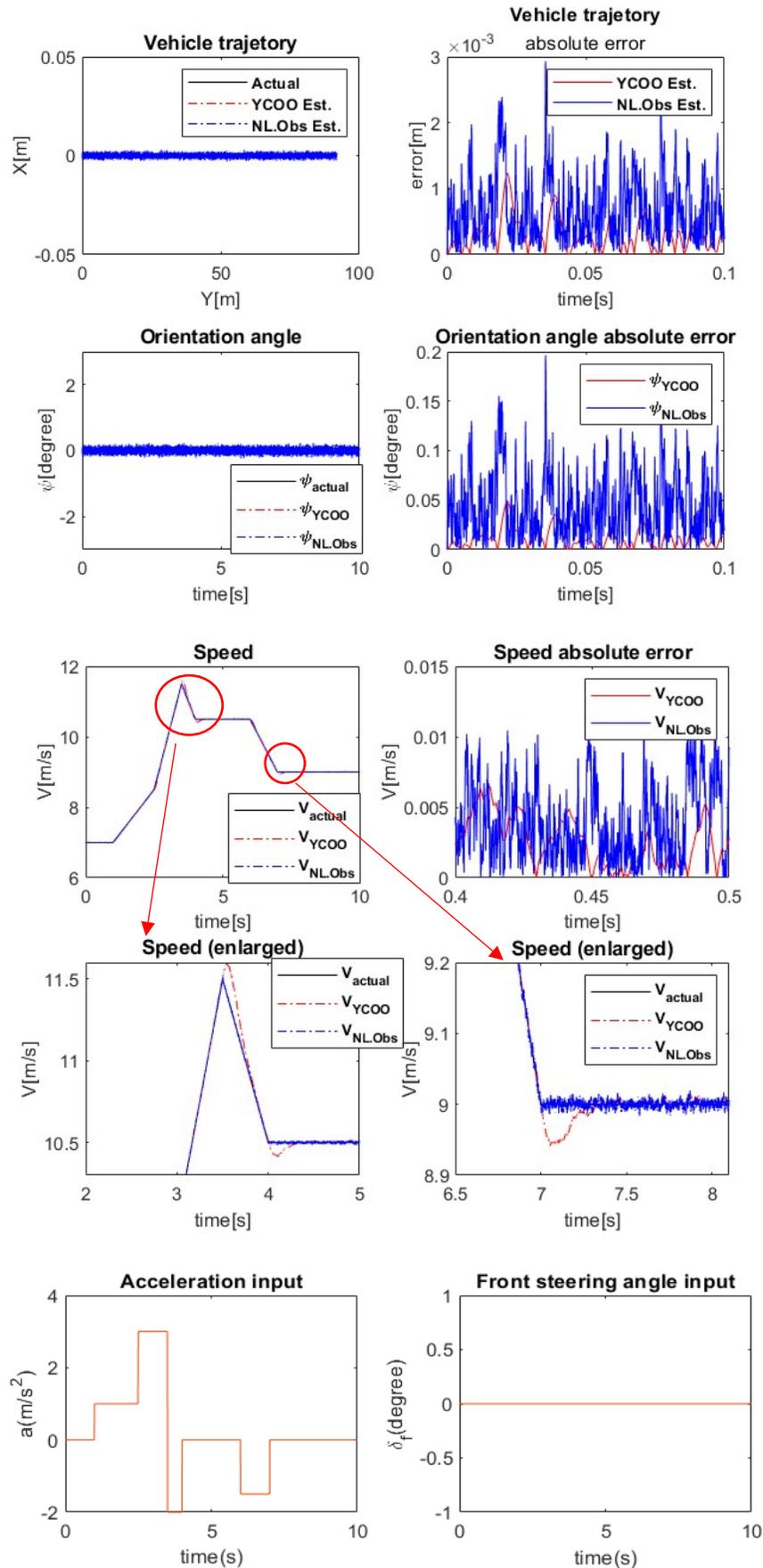

Figure 7. Simulation results: Straight-line driving.

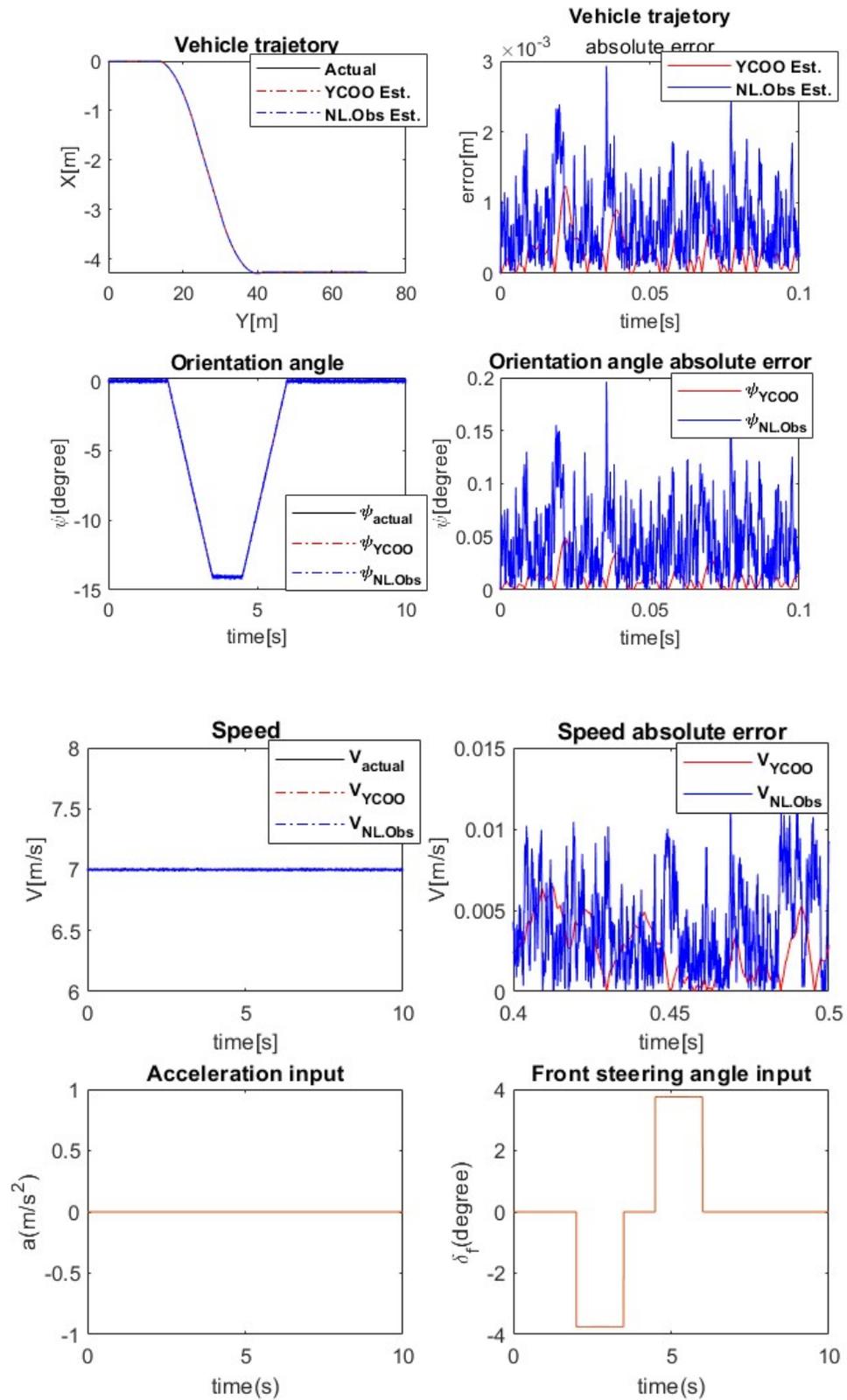

Figure 8. Simulation results: Lane change maneuver

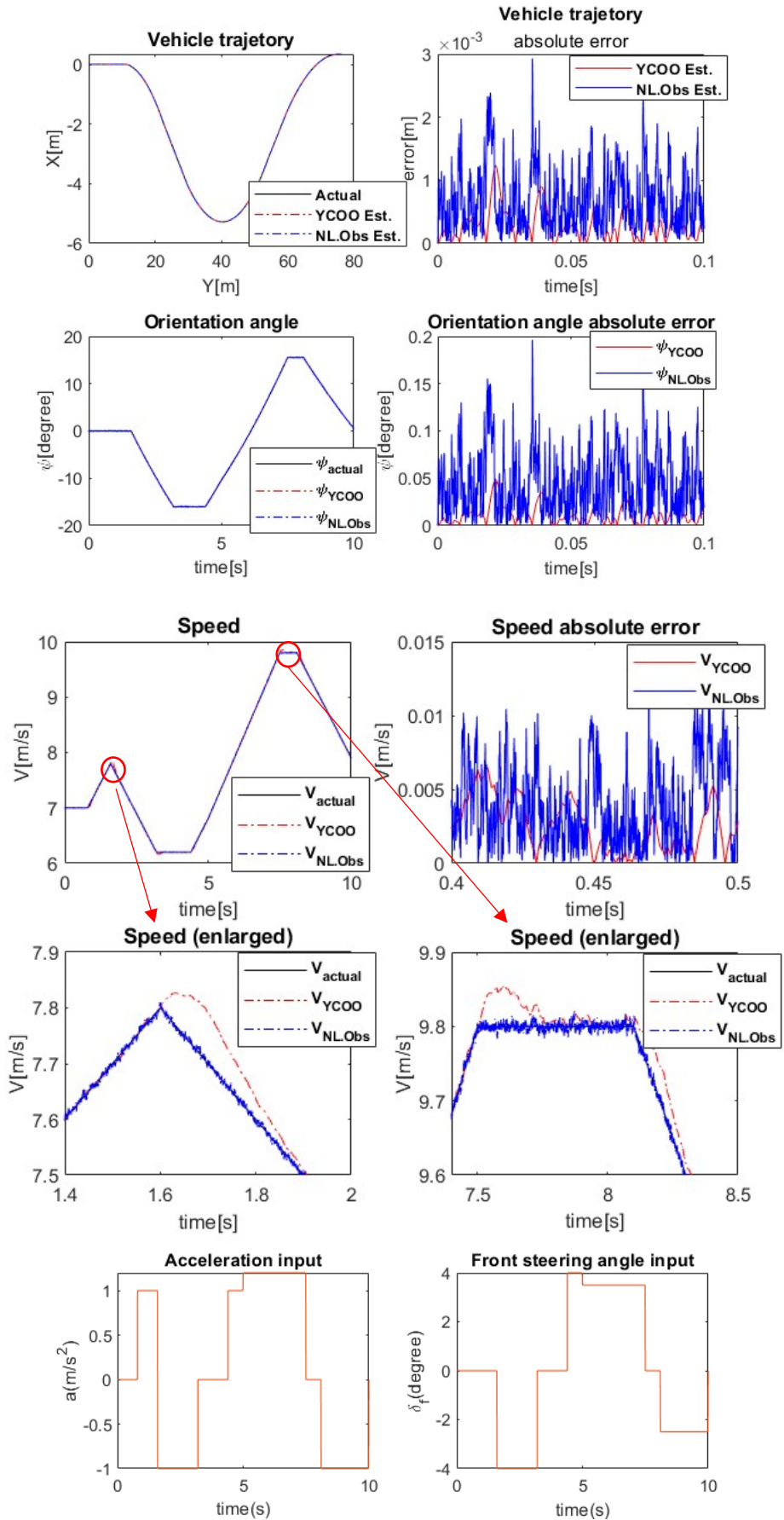

Figure 9. Simulation results: Double-lane change maneuver

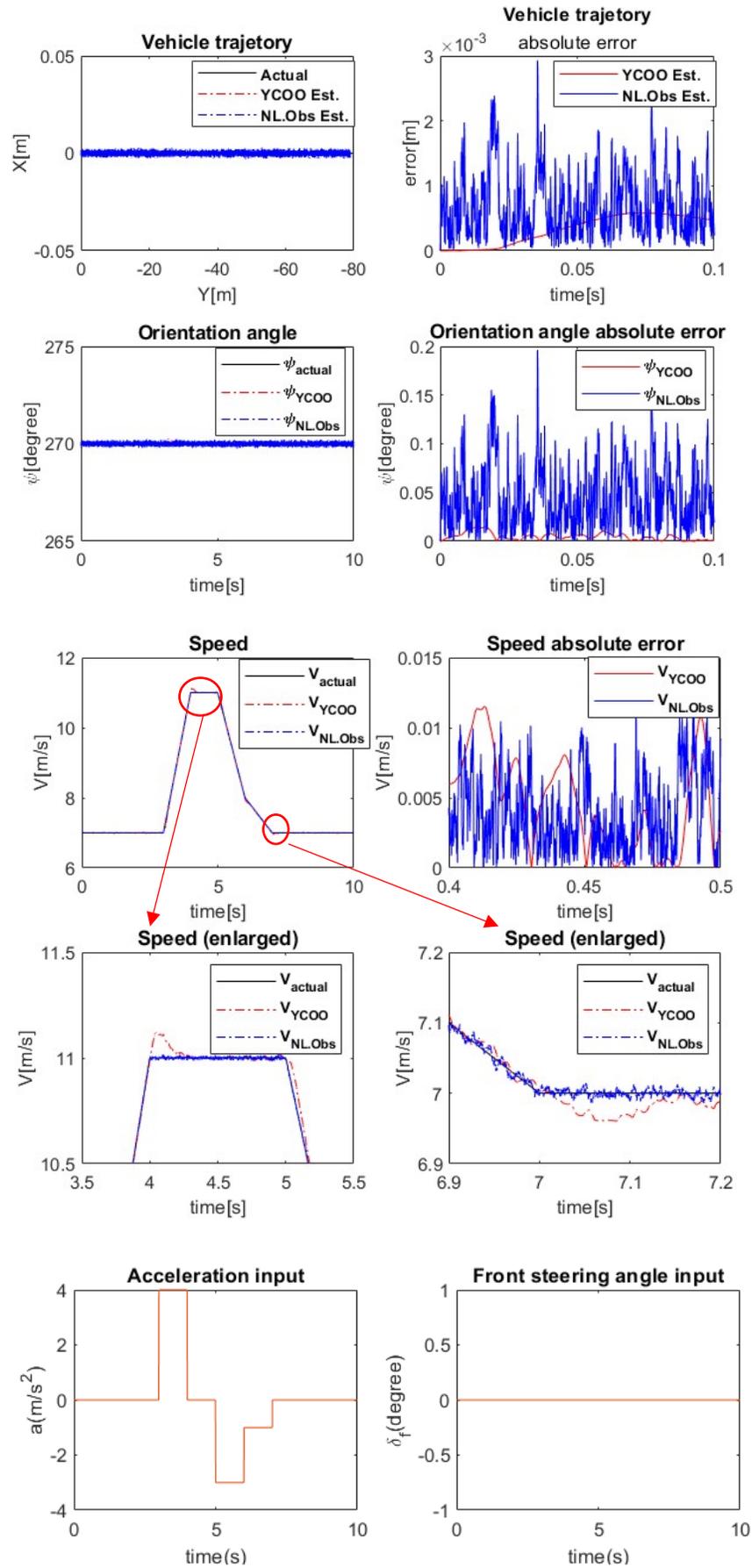

Figure 10. Simulation results: Cross traffic driving

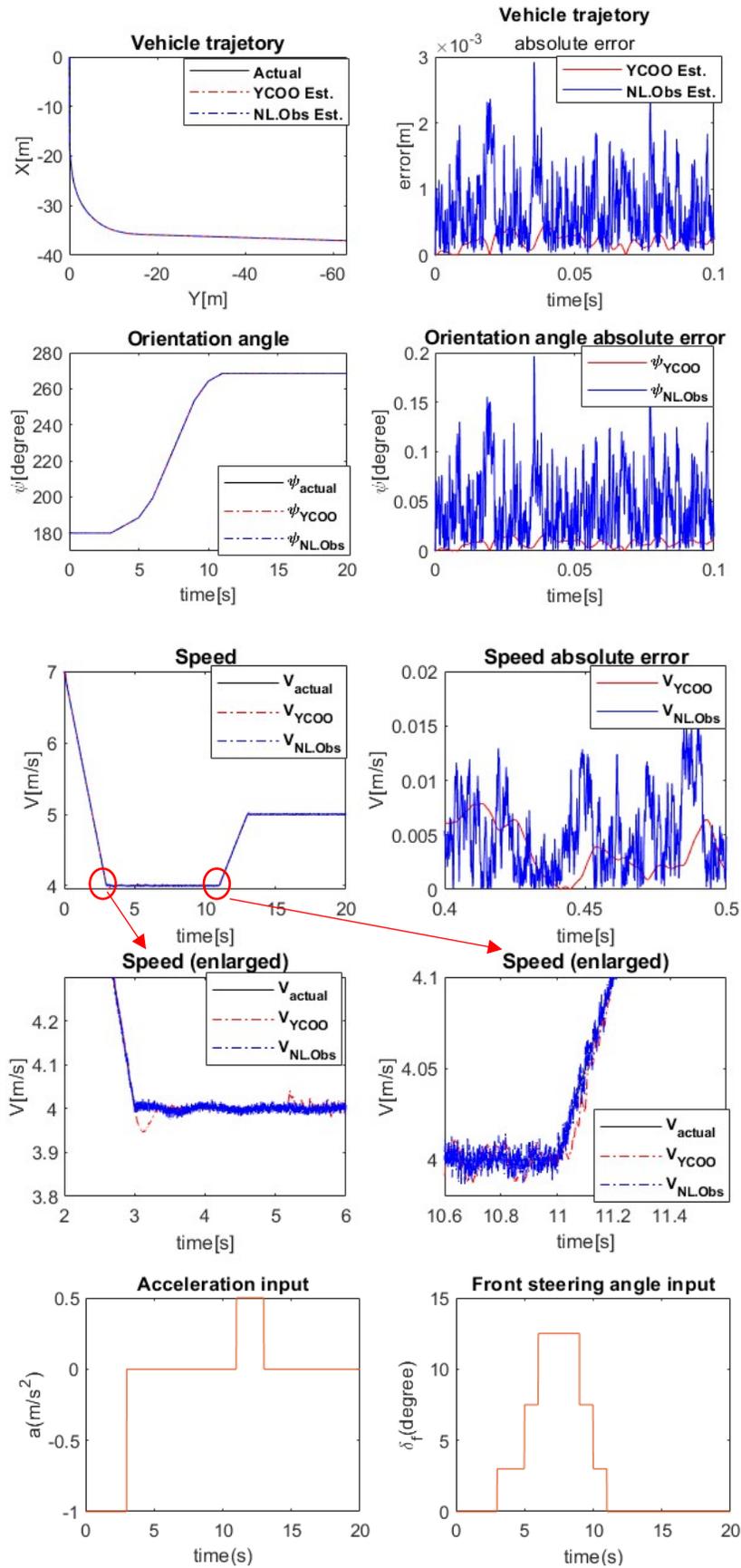

Figure 11. Simulation results: Left turn maneuver

TABLE V
. Comparison between YCOO and Nonlinear observer on average RMS of trajectory,
Orientation angle and speed estimations in five vehicle motion scenarios

| Vehicle motions | RMS of vehicle trajectory ($m$) | | RMS of orientation angle ($degree$) | | RMS of speed ($m/s$) | |
|---|---|---|---|---|---|---|
| | YCOO Est. | NL.Obs Est. | YCOO Est. | NL.Obs Est. | YCOO Est. | NL.Obs Est. |
| Fixed lane | 0.0001 | 0.0001 | 0.0209 | 0.0599 | 0.0048 | 0.0302 |
| Lane change | 0.0001 | 0.0002 | 0.0207 | 0.0599 | 0.0003 | 0.0005 |
| Double lane | 0.0001 | 0.0022 | 0.0277 | 0.0599 | 0.0048 | 0.0155 |
| Cross traffic | 0.0001 | 0.0080 | 0.0528 | 0.0599 | 0.0048 | 0.0223 |
| Left turn | 0.0001 | 0.0194 | 0.0466 | 0.0627 | 0.0051 | 0.0097 |

❖ Values highlighted in red indicate that the RMS errors exceed the set tolerance of 0.05.

TABLE VI
. Comparison between YCOO and Nonlinear observer on errors average standard deviation of trajectory,
Orientation angle and speed estimations in five vehicle motion scenarios

| Vehicle motions | STD of vehicle trajectory error ($m$) | | STD of orientation angle error ($degree$) | | STD of speed error ($m/s$) | |
|---|---|---|---|---|---|---|
| | YCOO Est. | NL.Obs Est. | YCOO Est. | NL.Obs Est. | YCOO Est. | NL.Obs Est. |
| Fixed lane | 0.0003 | 0.0005 | 0.0050 | 0.0209 | 0.0048 | 0.0300 |
| Lane change | 0.0005 | 0.0012 | 0.0005 | 0.0207 | 0.0036 | 0.0048 |
| Double lane | 0.0005 | 0.0013 | 0.0050 | 0.0270 | 0.0048 | 0.0151 |
| Cross traffic | 0.0002 | 0.0005 | 0.0126 | 0.0133 | 0.0121 | 0.0382 |
| Left turn | 0.0005 | 0.0073 | 0.0060 | 0.0420 | 0.0054 | 0.0108 |

TABLE VII
Statistical SIGNIFICANCE ANALYSIS: P-VALUE BETWEEN TWO METHODS IN ESTIMATION OF TRAJECTORY,
ORIENTATION ANGLE AND SPEED IN FIVE VEHICLE MOTION Scenarios

| Vehicle motions | p-value of vehicle trajectory estimation | p-value of orientation angle estimation | p-value of speed estimation |
|---|---|---|---|
| Fixed lane | 1 | $2.37\times 10^{-11}$ | $7.39\times 10^{-5}$ |
| Lane change | 0.6758 | $2.84\times 10^{-11}$ | 0.007 |
| Double lane | $5.92\times 10^{-10}$ | $3.71\times 10^{-7}$ | $7.43\times 10^{-4}$ |
| Cross traffic | 0 | 0.038 | 0.022 |
| Left turn | $8.22\times 10^{-15}$ | 0.046 | 0.043 |

❖ Values highlighted in red indicate that the p-value exceed the set significance level of 0.05.

TABLE VIII
. Comparison between YCOO and Nonlinear observer on Error Frequency of trajectory,
Orientation angle and speed estimations in five vehicle motion scenarios

| Vehicle motions | Error frequency of vehicle trajectory ($Hz$) | | Error frequency of orientation angle ($Hz$) | | Error frequency of speed ($Hz$) | |
|---|---|---|---|---|---|---|
| | YCOO Est. | NL.Obs Est. | YCOO Est. | NL.Obs Est. | YCOO Est. | NL.Obs Est. |
| Fixed lane | 4.393 | 30.021 | 6.191 | 42.630 | 0.166 | 28.113 |
| Lane change | 0.020 | 30.152 | 3.920 | 42.603 | 0.817 | 28.121 |
| Double lane | 0.333 | 30.045 | 2.593 | 42.596 | 0.215 | 28.098 |
| Cross traffic | 0.028 | 30.041 | 0.136 | 42.612 | 0.309 | 28.113 |
| Left turn | 0.004 | 30.367 | 0.046 | 38.738 | 0.671 | 24.987 |

TABLE IX
COMPARISON BETWEEN YCOO AND NONLINEAR OBSERVER ON ROBUSTNESS
AGAINST $l_t$ VARIATION OF ORIENTATION ANGLE AND SPEED ESTIMATIONS

| $l_t$ variation level | RMS of orientation angle (*degree*) | | RMS of velocity (*m/s*) | |
|---|---|---|---|---|
| | YCOO Est. | NL.Obs Est. | YCOO Est. | NL.Obs Est. |
| $l_t/l_{t\ norminal} = 1$ | 0.0001 | 0.0001 | 0.0001 | 0.0001 |
| $l_t/l_{t\ norminal} = 0.8$ | 0.001 | <span style="color:red">0.053</span> | 0.012 | <span style="color:red">0.143</span> |
| $l_t/l_{t\ norminal} = 1.2$ | 0.001 | <span style="color:red">0.133</span> | 0.011 | <span style="color:red">0.134</span> |

❖ Values highlighted in red indicate that the RMS errors exceed the set tolerance of 0.05.

## 5 Conclusion

This paper compares the performance of two observer systems developed from the same nonlinear vehicle tracking model. The first system, a nonlinear observation system based on Luenberger observers, utilizes four different gains to operate across the entire range. In contrast, the second system, the YCOO system, is implemented with only three observers derived from linearized models. Both systems provide stable and accurate estimates across the operating range. However, simulation results summarized in Table VII demonstrate that the YCOO system is significantly more effective at rejecting high-frequency noise with a confidence level over 95%. On average, it reduces the root mean square (RMS) and error frequencies by at least 2–3 times compared to the nonlinear system, as shown in Tables V and VIII.

Both systems demonstrate robustness against parameter variations, but the YCOO system shows superior performance, particularly when the vehicle's wheelbase is varied. For instance, with a 20% variation in wheelbase length, the YCOO system increases the magnitude of estimation errors only by approximately 10 times for orientation and 100 times for speed. In comparison, the nonlinear system exhibits a significantly larger increase—around 1000 times for both states. Despite experiencing more tracking loss during discontinuous state changes, the YCOO system is simpler to implement and requires less computational effort in real-world applications compared to the nonlinear Luenberger observer, which requires four different gains. Additionally, the YCOO system offers enhanced noise reduction and superior robustness to model variations, making it a promising and practical solution for improving real-world vehicle tracking systems. Its balance of computational efficiency, ease of implementation, and noise rejection capabilities positions it as a strong candidate for future applications in vehicle tracking and control systems.

Although experimental tests conducted in paper [1] demonstrated that the simple kinematic vehicle model described in Section 1.2 performs effectively in real urban traffic tracking tasks, a kinematic model is insufficient for high speed driving or complex road conditions. For example, at high speeds, tires generate lateral forces through slip angles which cannot be captured by a kinematic model. Also, mass and yaw inertial, which are not modeled in the kinematic model, heavily influence turning behavior in high speed. Therefore, a future research could focus on extending the application of the YCOO system to more complex dynamic vehicle tracking models. For instance, applying YCOO to a bicycle model could enhance its capability to estimate states in more challenging scenarios involving rapid changes or high-speed dynamics, such as pre-crash avoidance. In addition, future study may also focus on modeling of systemic delay occurred in the vehicle tracking system and testing the performance of YCOO along with the Smith Predictor. Those extensions would allow the YCOO system to address a broader range of real-world vehicle tracking situations and further validate its robustness and versatility.

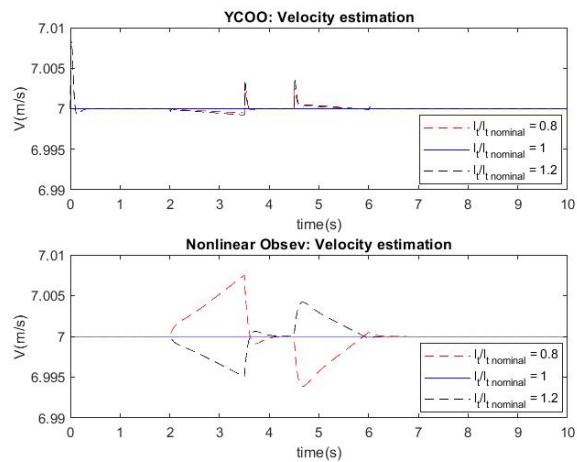

Figure 12. Robustness against $l_t$ variation: velocity estimation.

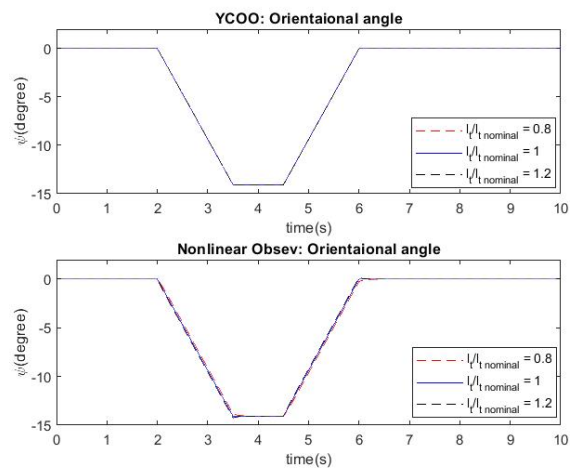

Figure 13. Robustness against $l_t$ variation: orientation angle estimation.